# Coronal electron density, temperature and solar spectral irradiance during the solar cycle 23 and 24

J.M Rodríiguez Gómez[1] and L. Vieira[1] and A. Dal Lago[1] and J. Palacios[2].

[1]National Institute for Space Research (INPE) and [2]Space Weather Group, Departamento de Fíisica y Matemáticas, Universidad de Alcalá University Campus.

**Abstract**

The plasma parameters such as the electron density and temperature plays a key role in the dynamics of the solar atmosphere. These characteristics are important in solar physics, because they can help to understand the physics in the solar corona. The goal is to reconstruct the electron density and temperature distributions in the solar corona. The relations between emission and plasma parameters in different time scales are studied. We present a physics-based model to reconstruct the density, temperature and emission in the EUV band. This model called CODET is composed of a flux transport model, an extrapolation model, an emission model and an optimization algorithm. The CODET model parameters were constrained by comparing the model's output to the TIMED/SEE record instead of direct observations because it covers a longer time interval than the direct solar observations currently available. The most important results of the current work that the recovery of SSI variability in specific wavelengths in the EUV band, also the variations in density and temperature in large time scale through the solar atmosphere, with the CODET model. The evolution of the electron density and temperature profiles through the solar corona in different layers during the solar cycle 23 and 24, will be presented. The emission maps were obtained and they are in accordance to the observations. Also, the density and temperature maps are related to the variations of the magnetic field in different layers through the solar atmosphere.

## 1 Introduction

The solar magnetic field and their relationship with the plasma parameters are important to describe some phenomena in the solar atmosphere. The magnetic field is created in the solar interior by the solar dynamo action (Dikpati and Gilman, 2009). This manifestation is observed as a variety of phenomena in the solar atmosphere (Mackay and Yeates, 2012; Solanki et al., 2006; Low, 1996; Hargreaves, 1995).

There are some processes where coronal electrons are accelerated and emit radiation, such as the acceleration by the electromagnetic field of photospheric radiation (i.e. EUV emission is produced by free-free emission from the chromosphere and corona). In general, the Solar Spectral Irradiance (SSI) influences the Earth's atmosphere for each wavelength in different altitudes. The EUV emission has considerable impact on the Earth's upper atmosphere, i.e., on the density, temperature, and total electron content, and it is an important driver for space weather (Schmidtke, 2015; Haberreiter et al., 2014). The study of plasma parameters such as electron density and temperature can contribute to understanding some phenomena shown in the solar atmosphere. Determinations of coronal densities have been made since ~

1950 from van de Hulst (1950) and Pottasch (1964, 1963) models from eclipse observations and empirical laws relating brightness with height. However, the measurement of these parameters is not trivial in the solar corona because the plasma is optically thin and the information received is integrated along the line of sight mixing information from different wavelengths (Kramar et al., 2014; Singh et al., 2002). For this reason, it is important to build models that can be used to study this behaviour and check whether or not the results are related to characteristics of the solar cycle and if they are changing in different time scales.

Density and temperature profile variation along the solar cycle is an important fact and gives clues for the solar corona dynamics. On the other hand, the problem of the heating corona is of great interest for solar physics. The primary conclusion is that the heating can be explained by processes that involve magnetic fields (Galsgaard and Nordlund, 1997). In this context, we decided to build a physics-based model that relies on the assumption that the density, temperature and emission variations are due to the evolution of the structure of the solar magnetic field. The COronal DEnsity and Temperature (CODET) model allows us to investigate some important aspects such as variations of density and temperature through the solar corona, in different heights and time scales. These variations are examined in large scale during the solar cycle 23 and 24. This model is based on the idea presented by C. Marqué and M. Kretzschmar in the poster entitled: Forward modeling of the electron density and temperature distribution in the corona using EUV and radio observations, in the LWS meeting, Boulder, 2007.

We structure the paper as follows. In Sect. 2, we describe the physics-based model, the Coronal DEnsity and Temperature (CODET) model. In Sect. 3, the main results are presented: reconstructions of Solar Spectral Irradiance (SSI) variability at 19.3 and 21.1nm, the density and temperature profiles during the solar cycle 23 and 24, the density and temperature maps through different layers in the solar atmosphere and the emission maps in different layers at 19.3nm and 21.1nm. In the Sect. 4, the discussion is presented. Finally, the concluding remarks are in Sect. 5.

## 2 The COronal DEnsity and Temperature (CODET) model

The COronal DEnsity and Temperature (CODET) model (Figure 1) uses a flux transport model of Schrijver (2001). The flux transport model is a key component of the proposed model. The flux transport model employed line-of-sight magnetic field data from SOHO/MDI and SDO/HMI full-disk magnetograms. These data are assimilated into the flux transport model to describe the dynamics of the solar photosphere. In this approach we could not employ directly the magnetic field strength density from MDI/SOHO (Scherrer et al., 1995) and HMI/SDO (Scherrer et al., 2012). The main reason is due to the synoptic maps do not take into account correctly the evolution of the active regions as they transit in the far-side. Additionally, synoptic maps do not cover most of the time the evolution of poles. The evolving surface-flux assimilation model is sampled every six hours from 1 July 1996. The assimilation model assumes that the magnetic field from SOHO/MDI and SDO/HMI is strictly vertical and the magnetograms are incorporated within 60° from disk center. The assimilation procedure is a straightforward mapping: after re-binning to a resolution of 8 arcsec, each magnetogram pixel is assumed to correspond to a single concentration at the corresponding latitude and longitude. In addition to the assimilated magnetograms, small magnetic bipoles (|φ| < 2 × 10 20 Mx) are injected outside the assimilation area.
This maintains the quiet-Sun network, which impacts the flux dispersal even though it adds little to the large-scale coronal field. Also, bipoles are inserted on the far-side of the Sun depending on the pattern and magnitude of the measured travel time differences of p-modes reflecting around the antipode of disk center (Schrijver and De Rosa, 2003).

These data are then used as boundary conditions for a series of potential-field source-surface (PFSS) extrapolations. The structure of the coronal magnetic field is estimated employing the Potential Field Source Surface (Schrijver, 2001; Schrijver and De Rosa, 2003). The Potential Field Source Surface (PFSS) model, extrapolates the line-of-sight surface magnetic field through the corona with the boundary assumed to be at the source surface and assuming that the solar corona is current free. The magnetic field is extrapolated from the photosphere at 1R to the corona at 2.5R.
The flux transport model is from the scheme of Schrijver (2001). This model considered fickian diffusion in all scales. The flux injection is described by a combination of random processes, capturing the properties of flux evolution (the flux emergence from the interior, flux dispersal over the surface and the flux disappearance from the photosphere). After the flux in a bipolar region has fully emerged, the region decays and the flux disperses across the surface. The flux dispersal in the photosphere is frequently modelled as a passive random walk diffusion, involving supergranulation, meridional flow and differential rotation.
The bipolar region source function is:

$$n(S, A)dSdt = (a_0 A S^{-p} + a_1 A^\alpha S^{-p-1})dSdt \quad (1)$$

where A[M x] is the flux injection parameter related to different levels of activity, S[deg 2 ≈ 150M m 2 ] is the area of the bipolar regions. At solar cycle maxima, the coefficients a 0 = 8 and p = 1.9 are determined by a fit to the area distribution for emerging active regions as derived by Zwaan and Harvey (1994), a 1 is the set to 8 deg −2 day −1 hemisphere −1 in order to match the total flux input. The weaker cycle dependence for the ephemeral region frequency compared to the active region frequency is approximated through a power-law scaling with the flux emergence parameter A with a power law index α = 1/3 (Schrijver, 2001).

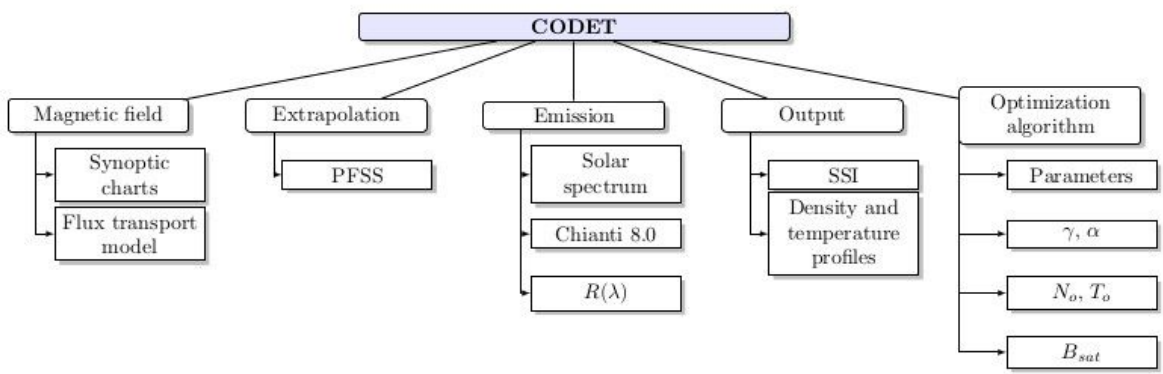

Figure 1: Schematic description of the COronal DEnsity and Temperature (CODET) model.

Also, we use an emission model. This model is based on the CHIANTI atomic database 8.0 (Del Zanna et al., 2015) using specific lines (19.3nm and 21.1nm) in the EUV band. This model considers the coronal abundances and ionization equilibrium to build the solar spectrum and modelling the electron density and temperature through the solar corona. Additionally, an optimization algorithm was used. The optimization algorithm Pikaia is a method for optimization based on a genetic algorithm (Charbonneau, 1995). Pikaia algorithm was used to determine some parameters in different problems such as the study of solar phenomena. It was used for empirical modelling of the solar corona (Gibson and Charbonneau, 1998), Doppler shifts of solar ultraviolet emission lines (Peter and Judge, 1999), and modeling the evolution of the solar irradiance (Vieira et al., 2011; Vieira and Solanki, 2010; Krivova et al., 2010, 2007). In this case, the Pikaia algorithm was used to search the best fit parameters of the CODET model.

## 2.1 Approach

The density and temperature profiles are related to the magnetic field. It is considered the thin flux tube model (Solanki, 1993); the magnetic field is bundled into discrete elements of concentrated flux, called frequently magnetic flux tubes. These tubes cover fractions of the solar surface (Fligge and Solanki, 2000). The flux tubes are narrow, the inflow of radiation through the hot walls exceeds the energy blocked. The geometry of the small-scale fields causes a non isotropic radiation field. The combination of these effects leads to variations in the solar irradiance on time-scales from days, to years (Vieira et al., 2012). A pressure balance is considered between the tube and the ambient (Vekstein and Katsukawa, 2000).

$$\underbrace{-\nabla \left( p + \frac{B^2}{8\pi} \right)}_{The\ gradient\ of\ total\ pressure} + \underbrace{\frac{1}{4\pi} (\mathbf{B} \cdot \nabla) \mathbf{B}}_{The\ magnetic\ tension} = 0 \qquad (2)$$

The vertical flux tubes are assumed not to be curved and thus do not have magnetic tension (neglected the second term of the Equation 2). Then, the pressure balance requirement is

$$2Nk_BT = \frac{B^2}{8\pi} \qquad (3)$$

where N [cm$^{-3}$] is the electron density, $k_B$ [erg K$^{-1}$] is the Boltzmann constant, T [K] is the temperature and B[G] is the magnetic field. Also, the plasma β (≈ 8πN $k_B$ T /B$^2$) from the Equation 3, is assumed to be small enough for the plasma to be effectively confined by the magnetic field (Emslie and Brown, 1980).
Then, considering the magnetic field:

$$B(r, \theta, \phi) = \sqrt{((B_r(r, \theta, \phi))^2 + (B_\phi(r, \theta, \phi))^2 + (B_\theta(r, \theta, \phi))^2)} \qquad (4)$$

where $B_r(r, \theta, \varphi)$, $B_\varphi(r, \theta, \varphi)$ and $B_\theta(r, \theta, \varphi)$ are the magnetic field components from Potential Field Source Surface (PFSS). The magnetic field B is measured in [G] units. We consider $B(r, \theta, \varphi) = B$ in the following description.

In this approach we use scaling laws for coronal loops in hydrostatic energy balance (Rosner-Tucker-Vaiana (RTV) scaling law). Scaling laws provide important diagnostics and predictions for specific physical models of the solar corona. These models have been widely applied in plasma physics, astrophysics, geophysics, and the biological science (Aschwanden et al., 2008). Scaling laws can be derived both from observations and theory, and the results can be described some characteristics and phenomena in the solar corona. We follow the simplest rule, the dependence on the squared electron density, which is also proportional to the optically thin emission measure in EUV, and thus to the observed flux (Aschwanden, 2005). In general, the RTV scaling laws express an energy balance, using approximations of constant pressure (Equation 3), no gravity and uniform heating. A special case of scaling law is related to magnetic scaling.

Here we employ the density and temperature distribution in function of the magnetic field. This dependency is employed by several authors, for example: Robbrecht et al. (2010); Vekstein and Katsukawa (2000); Yokoyama and Shibata (2001); Golub et al. (1980); Golub (1983); Emslie (1985); Brown et al. (1979). These authors use scaling laws to describe the density and temperature profiles in function of the magnetic field. These models take into account also other parameters such as the flux tube loop length (L), volume (V) and heat conditions (τ). Mandrini et al. (2000) discuss in details the scaling laws employed by different models of coronal heating and their relation to the magnetic field. In our model, we decided to consider the dependency just on the magnetic field intensity, that is, the model exponents for the loop length (L) and volume (V) are considered equal to zero. The main reason for this assumption is related to the number of free parameters needed in the optimization algorithm and the time needed to parametrize each flux tube. Also, it is assumed that the total plasma pressure remains unchanged in each flux tube (Equation 3). Then, an analytic treatment is possible, thus highlighting the essential physics of the simplified problem and allowing us to develop the simple scaling laws (Emslie, 1985). We employ the distribution of density and temperature in the following way:

$$N(B) = N_o \left(\frac{B}{B_s}\right)^\gamma \quad [cm^{-3}] \quad (5)$$

We consider the function $B_f(R)$:

$$B_f(R) = b_{f0} \times e^{-\left(\frac{R}{\tau_{bf}}\right)^2} \quad [G] \quad (6)$$

where $b_{f0}$ [G] units and $\tau_{bf}$ [R] units are constant values (in this case we use $b_{f0}$ = 20G and $\tau_{bf}$ = 1.2R ); R corresponds to the height through the solar atmosphere and it varies from 1R to 2.5R. It was defined to describe two different temperature regimes related to regions with strong or weak photospheric magnetic field, using the following conditions:

$$\text{if } B < B_f(R)$$
$$T(B) = T_o \quad [K] \tag{7}$$

$$\text{If } B > B_f(R)$$
$$T(B) = T_o \left(\frac{B}{B_s}\right)^\alpha \quad [K] \tag{8}$$

where γ and α are power law indices, $(B/B_s)$ is the factor related to the amount of flux in each pixel, $B_s$ [G] is a constant value of the magnetic field, $N_o$ [cm$^{-3}$] and $T_o$ [K] are background density and temperature. The temperature T and electron density N are measured in [K] and [cm$^{-3}$] respectively. Additionally, in this approach was evaluated the exponent value of the scaling law in B. In addition temperature considerations between open and closed field lines are defined in Equations 7 and 8.

**2.2 Emission measure formalism**

Different models were employed to describe the emission measurement in different wavelengths (Kretzschmar et al., 2006; Warren, 2006; Warren et al., 1998; Vernazza et al., 1981). In this section, some characteristics of emission measure formalism used in the CODET model will be described. Assuming that the emission lines are optically thin, it is possible to measure only the integrated emission along a given line of sight, but it is necessary to consider the ionization and recombination coefficients related to the contribution function. This emission line depends on the atomic transitions and the conditions of the solar atmosphere. The specific intensity can be described by:

$$I_o(\lambda) = \int \int R(\lambda) G(\lambda, T) \, d\lambda \, N^2 ds \tag{9}$$

where $G(\lambda, T)$ [erg cm$^3$ s$^{-1}$ sr$^{-1}$] is the contribution function from the CHIANTI atomic database 8.0, dλ [nm] is the differential element in wavelength, N [cm$^{-3}$] is the electron density, ds [cm] is the differential distance along the line-of-sight and R(λ) is the instrumental response. The contribution function was used to construct the solar spectra for a specific wavelength. This function contains relevant atomic physical parameters such as ionization equilibrium and coronal abundances. The ionization equilibrium from Mazzotta et al. (1998) and coronal abundances from Meyer (1985) were used. These models are considered the solar corona as optically thin. Some of these contribution functions are shown in Rodríguez Gómez (2017) and Rodríguez Gómez et al. (2017); whereas the instrumental response depends on wavelength and temperature, it constitutes an important specification of the instruments; in this case we consider R(λ) = 1 (ideal case). The intensity I is the full-disc average intensity measured at Earth from an emission line, where D = 1AU = 1.4960 × 10$^{11}$ m.

$$I = \frac{I_o}{D^2} \quad [W/m^2/nm] \tag{10}$$

**2.3 Optimization Algorithm**

The optimization algorithm was used to search the best fit parameters of the CODET model. In order to implement Pikaia Algorithm, we use BELUGA, which is a MATLAB optimization package and is freely available from Medical School at University of Michigan in the virtual physiological Rat Project. Beluga finds in a local minimum x of an objective function an initial population of candidate solutions. The free parameters are defined following:

$$par = par_{min} + (par_{max} - par_{min}) \times par(n) \tag{11}$$

where par is the free parameter that will be optimized by Pikaia algorithm, par max and par min are the lower and upper limits of parameters, par(n) should be located at the interval [0, 1], n is the number of free parameters. It is calculated a goodness-of-fit $\chi^2$ between TIMED/SEE and modelled data, in general $\chi^2 \leq 1$ indicates an acceptable fit. The goodness-of-fit is the key point between the Pikaia algorithm and the model of plasma parameters (Figure 2).

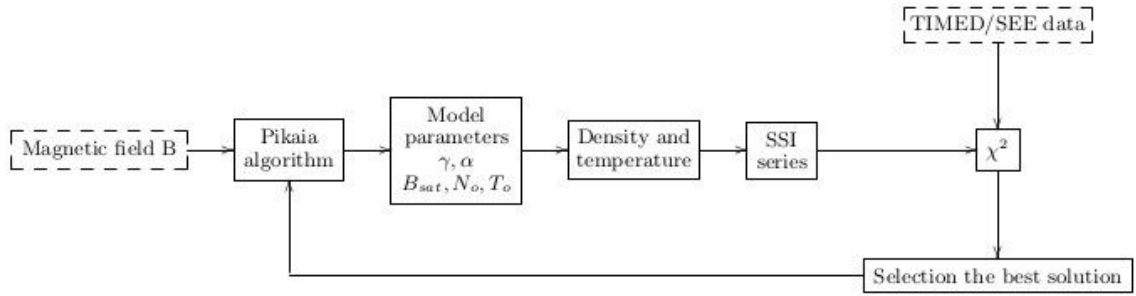

Figure 2: Schematic description of optimization algorithm Pikaia, where dashed boxes describe the input parameters.

The optimization algorithm was applied to fit two wavelengths 19.3nm and 21.1nm. The model parameters γ, α, $B_s$, $N_o$ and $T_o$ were adjusted. Several cases were explored to search the best fit between TIMED/SEE data and data from the CODET model. The $\chi^2$ function was defined after several tests as:

$$\chi^2 = \frac{(I_{model} - I_{obs})^2}{|I_{obs}|} \tag{12}$$

where $I_{model}$ is the intensity from our model and $I_{obs}$ corresponds to the intensity of TIMED/SEE data. In this case, it was chosen a period of ten days during the solar cycle 23 and 24 (Feb. 01 (2003), Oct. 01 (2003), Oct. 01 (2004), Oct.01 (2005), Oct. 01 (2007) Oct. 01 (2008), Oct. 01 (2009), Oct. 01 (2011), Oct. 01 (2014), Oct. 01 (2016) at 12 : 00UT). The characteristics evaluated in each case were:
1) Goodness-of-fit between Solar Spectral Irradiance from TIMED/SEE and modelled data.
2) Electron density and temperature profiles according to observational and model descriptions (Fontenla et al., 2014; Habbal et al., 2010; Golub and Pasachoff, 2009; Aschwanden, 2005; Withbroe, 1988; Billings, 1966).

The Solar Spectral Irradiance data used in this work is the TIMED/SEE from the NASA TIMED mission's Solar EUV Experiment (SEE) EUV Grating Spectrograph (EGS) merged with a model driven by The SORCE XUV Photometer System (XPS). The Model uses GOES XRS measurement data and CHIANTI spectral models as well. The CHIANTI spectral model includes the differential emission measures (DEMs) and also isothermal spectra appropriate for the Sun. It has been developed to process the measurements from broadband photometers. They are combined to match the signals from the XPS and produce spectra from 0.1 to 40 nm in 0.1 nm intervals (Woods et al., 2008, 2005; Woods and Rottman, 2005). Table 1 lists the parameters employed in Equations 5 and 8 that are used to compute the Solar Spectral Irradiance (Equation 10). In general the parameter values correspond to γ > 0, α < 0 and $B_s$ ≤ 10G.

Table 1: CODET model parameters: γ, α, $N_o$, $T_o$ and $B_s$. Typical values for best fits and specifications about the optimization algorithm: $\chi^2$, population size and generation.

| Parameter | Value | Units |
|---|---|---|
| $\gamma$ | 2.4459 | |
| $\alpha$ | −1.8502 | |
| $N_o$ | $2.3144 \times 10^8$ | $[cm^{-3}]$ |
| $T_o$ | $1.6093 \times 10^6$ | $[K]$ |
| $B_s$ | 4.4080 | $[G]$ |
| $\chi^2$ | 0.0017 | |
| Population size | 20 | |
| Generation | 70 | |

**3 Results**
**3.1 Solar Spectral Irradiance**

Solar Spectral Irradiance at Extreme UltraViolet wavelengths (EUV) drives physical and chemical processes. The EUV solar irradiance is the most important parameter to monitor the space weather. The SSI variation at EUV wavelengths has important consequences for the Earth's upper atmosphere because the SSI is completely absorbed into the tenuous layers above the stratosphere (Fontenla et al., 2017; Scholl et al., 2016). Several models of EUV have been developed since 1970s, based on the reference irradiance spectrum and its extrapolation using proxies and using the different features of the solar atmosphere (Scholl et al., 2016; Thuillier et al., 2014; Ermolli et al., 2013; Kretzschmar et al.,2004). The emission at EUV wavelengths is originated in the solar chromosphere, transition region and corona; and it is affected by the Sun's magnetic field dynamics (Aschwanden, 2005; Warren et al., 1998). The SSI varies on timescales of minutes to days, the 27-day solar rotation period (manifested into synoptic charts using that input in this approach), and the 11-year solar cycle (this variation was analyzed in this article). The CODET model has as an output the Solar Spectral Irradiance from the photospheric magnetic field evolution over the solar cycle 23 and 24 (Figure 3 (a)) computed for three lines simultaneously (17.1nm, 19.3nm and 21.1nm), but the best results were obtained in two wavelengths 19.3nm and 21.1nm.

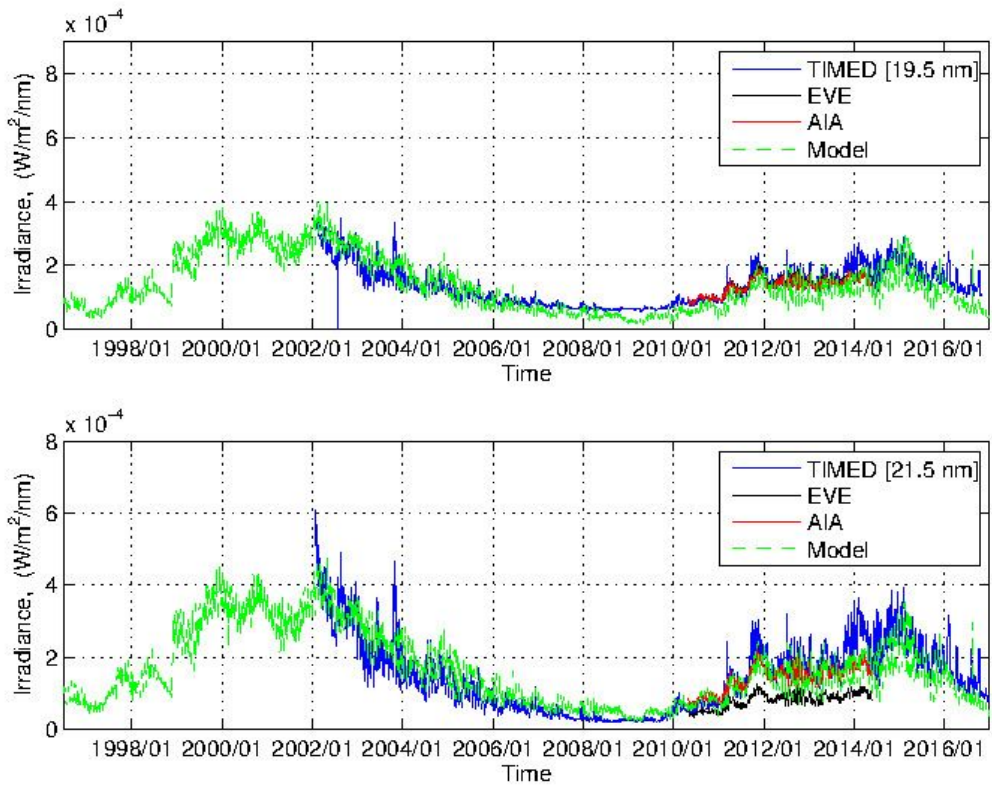

(a) SSI at 19.3nm and 21.1nm

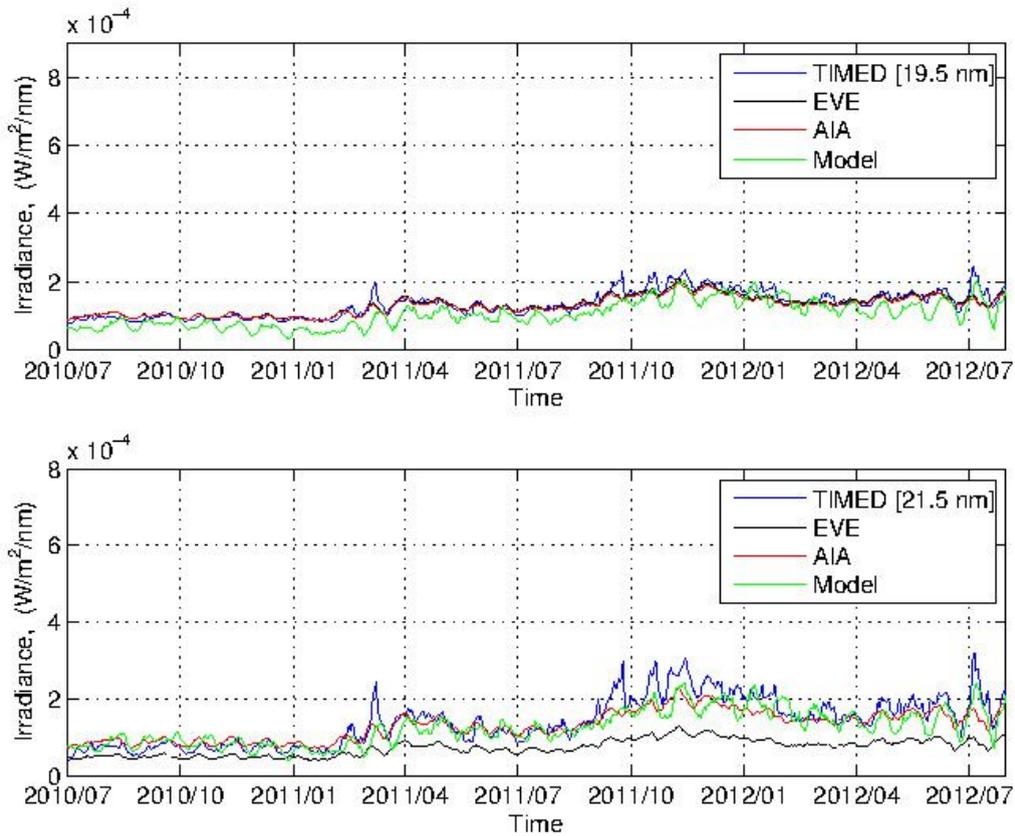

(b) Close-up/detail of SSI interval at 19.3 nm and 21.1 nm

Figure 3: Solar Spectral Irradiance (SSI) using the CODET model (green line) and Solar Spectral Irradiance from TIMED/SEE (blue line), SDO/EVE (black line) and SDO/AIA (red line). (a) SSI at 19.3 nm and 21.1 nm during the solar cycle 23 and 24. (b) The best fit interval of Solar Spectral Irradiance (SSI) from CODET model from Jul. 01 (2010) to Jul. 31 (2012).

The scatter plots were obtained in each case and the chi-squared test ($\chi^2$) was calculated. The chi-squared test was obtained for EVE/SDO ($\chi^2$ = 0.0745 at 19.3 nm and $\chi^2$ = 0.1534 at 21.5 nm) and AIA/SDO ($\chi^2$ = 0.8131 at 19.3 nm and $\chi^2$ = 0.2994 at 21.5 nm) data to review the consistency of the modelled values in comparison with observed data. The $\chi^2$ values from TIMED and EVE are very similar ($\chi^2$ = 0.0889 at 19.3 nm and $\chi^2$ = 0.0826 at 21.5 nm); this fact allow us to declare that the model parameters describe adequately observational data and their variations in long and short time scales. Two branch trends can be noticed: one of them overestimates observations slightly while other underestimates observations. The specific interval from Jul. 01 (2010) to Jul. 31 (2012) was selected to highlight the variations in short temporal scale. The variability was recovered and follow the observational data trend from EVE and AIA datasets (Figure 3 (b)). However, the EVE data are lower than CODET, TIMED and AIA data at 21.5 nm. In general, the Solar Spectral Irradiance from CODET model is consistent with daily variations from the instruments on board of SDO spacecraft. On the other hand, the correlation coefficient (R) analysis shows a strong linear relationship between modelled and TIMED data in both wavelengths (R = 0.750 at 19.3 nm and R = 0.796 at 21.5 nm). Also R value between EVE and CODET data shows a medium-weak relationship (R = 0.518 at 19.3 nm and R = 0.523 at 21.5 nm); while the correlation coefficient shows a weak relationship between the AIA and modelled data (R = 0.332 at 19.3 nm and R = 0.332 at 21.5 nm).

### 3.2 Density and temperature profiles during the solar cycle 23 and 24

Due to the problem with the direct measurements of the plasma parameters, profiles of electron density and temperature from the CODET model (Section 2.1) will be presented in this section. The plasma parameters, more specifically, the electron density and temperature shows an interesting behaviour through the solar atmosphere. The transition region demarcates the boundary where the chromospheric temperature increases and the density drops, also through the solar corona the temperature increases (from ~ $1 \times 10^5$ K to ~ $1 \times 10^6$ K) while the density decreases (from $1 \times 10^{10}$ cm$^{-3}$ to $1 \times 10^8$ cm$^{-3}$). These trends were reconstructed using the CODET model (Figure 4).

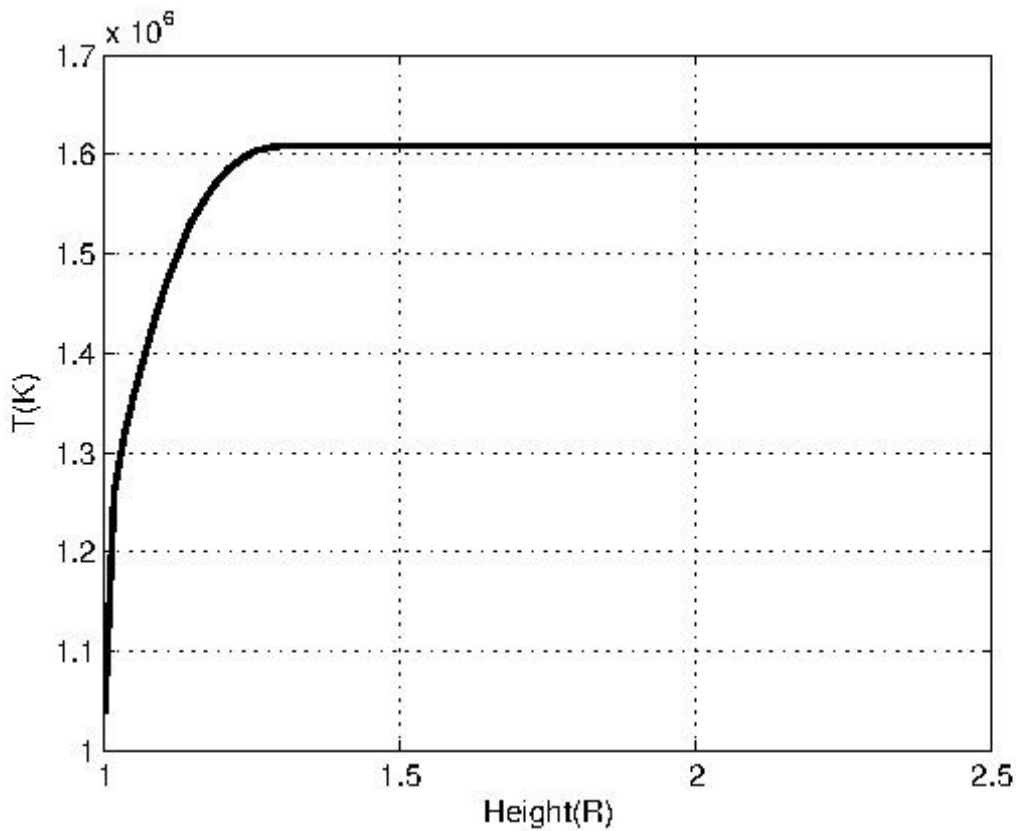

(a) Temperature profile

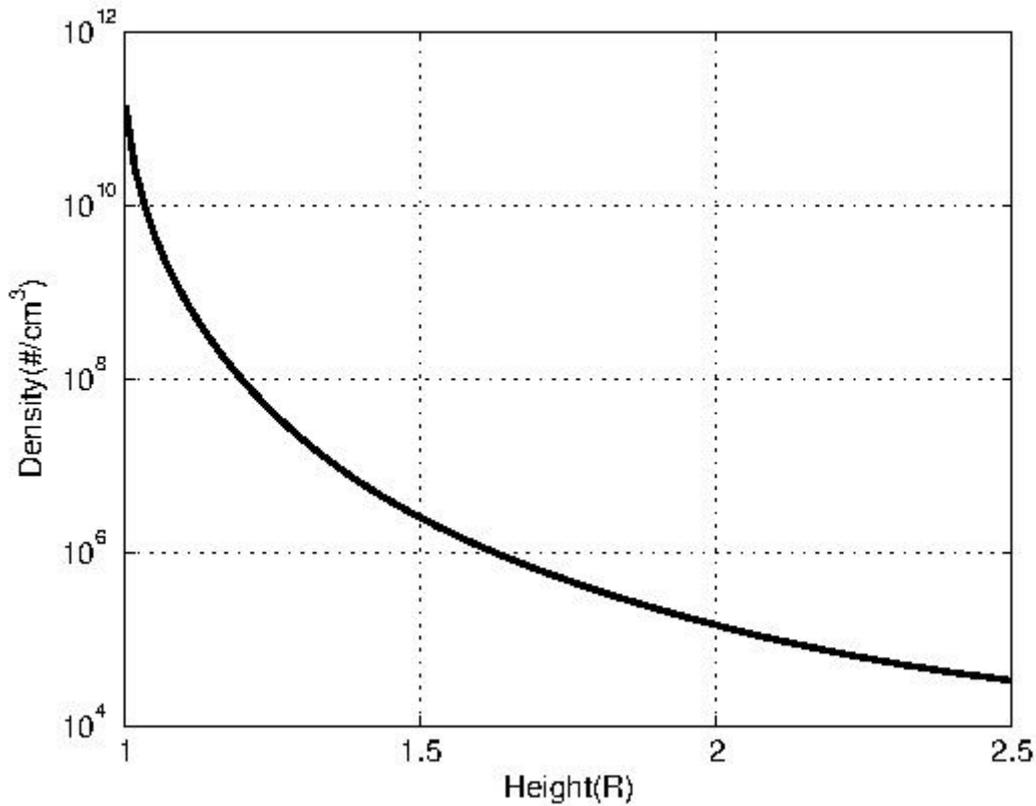

(b) Density profile

Figure 4: Temperature profile (left panel) and density profile (right panel) through the solar atmosphere on Dec. 15 (2001); specifically from R = 1R to R = 2.5R , for the parameter combinations shown in Table 1.

The evolution of the electron density and temperature profiles were obtained using Equations 5 and 8 and the parameters of the model were shown in Table 1. The electron density and temperature average profiles were obtained from different layers through the solar atmosphere R=1.14, 1.19, 1.23, 1.28, 1.34, 1.40, 1.46, 1.53, 1.61, 1.74, 1.79, 1.84 and 1.90 R . Variations in temperature and density during the last solar cycles are displayed in Figure 5. Lower values in temperature are shown in the solar cycle 23, while in the solar cycle 24 the temperature increases. The density values are higher in the solar cycle 23 and lower in the solar cycle 24. The external layers show lower values in average density than the layers near the photosphere.

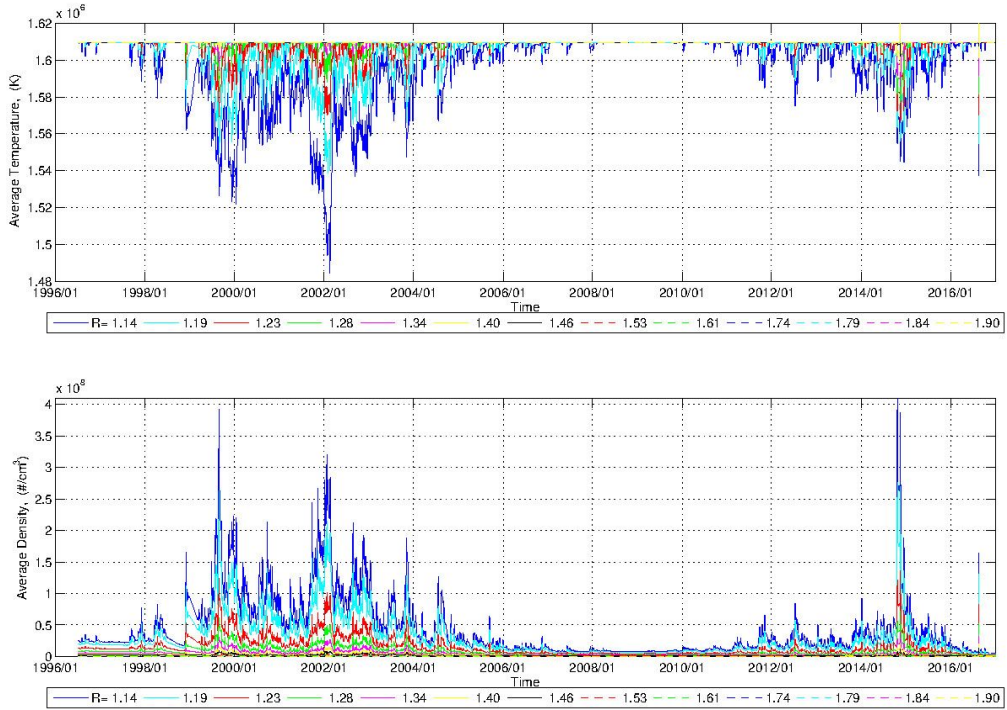

Figure 5: Average temperature (upper panel) and density (lower panel) profiles from CODET model, through different layers: R = 1.14, 1.19, 1.23, 1.28, 1.34, 1.40, 1.46, 1.53, 1.61, 1.74, 1.79, 1.84 and 1.90 R , during the solar cycle 23 and 24.

### 3.3 Density and temperature maps

Density and temperature maps in two different layers (R=1.16 and 1.30R) in the solar atmosphere are shown in Figure 6. The magnitude of the magnetic field was obtained from PFSS model from Equation 4 in these specific layers.

For these purposes three days were selected: Dec. 15 (2001), Nov. 15 (2003) and Feb. 05 (2015). Figure 6 shows regions with higher values in density in regions with stronger magnetic field. These regions are related to ARs located in the active region belts. The temperature maps show some structures with medium and lower values when the magnetic field is more intense. The temperature maps show the temperature variations from $\log_{10}T = 5.4$ to $\log_{10}T = 6.2$, while the density maps show variations from $\log_{10} N = 6.0$ to $\log_{10} N = 9.0$.

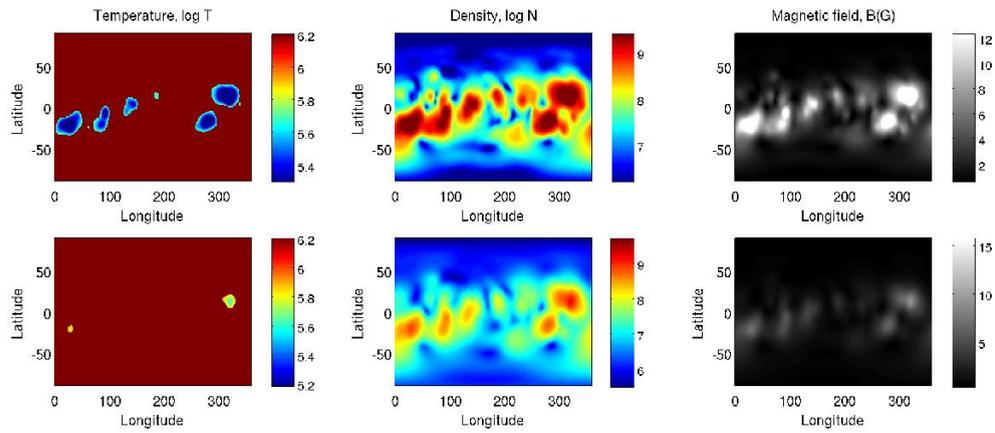

(a) Temperature, density and magnetic field maps in Dec. 15 (2001)

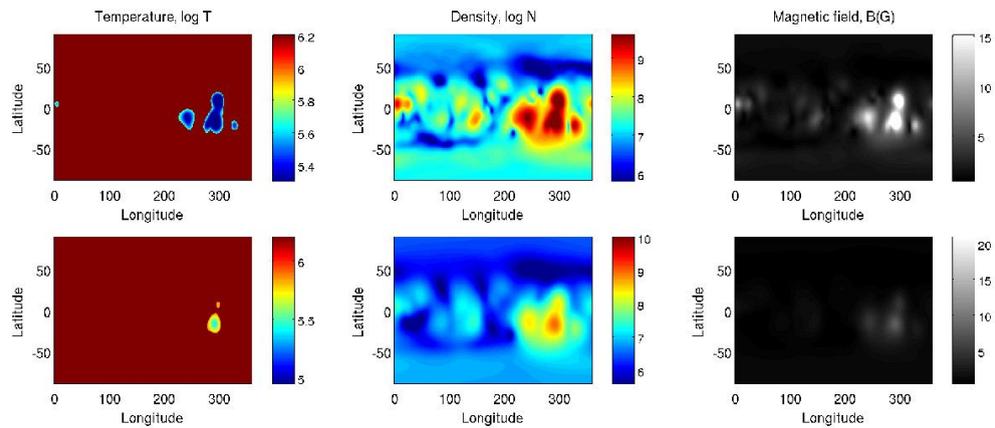

(b) Temperature, density and magnetic field maps in Nov. 15 (2003)

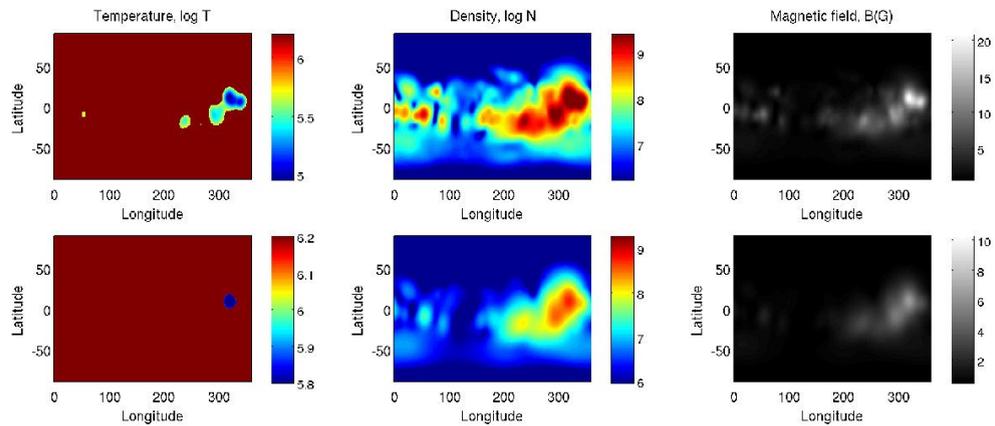

(c) Temperature, density and magnetic field maps in Feb. 05 (2015)

Figure 6: Comparison between temperature, density and magnitude of magnetic field in different layers first row R = 1.16R and second row R = 1.30R , using the CODET model. First column: temperature maps ($\log_{10}T$), second column: density maps ($\log_{10}N$) and last column: magnetic field maps (B(G)), a) Dec. 15 (2001), b) Nov. 15 (2003) and c) Feb. 05 (2015). All plots show different scales to highlight the characteristics in each day and layer.

## 3.4 Emission maps

The emission maps were obtained using two wavelengths: 19.3nm and 21.1nm in three different layers of the solar atmosphere R = 1.33, 1.42 and 1.60 R , in specific days: Dec. 15 (2001), Nov. 15 (2003) and Feb. 05 (2015) were explored. Figures 7 and 8 show the intensity maps in the transition region and the solar corona. The emission maps show regions with higher values in intensity over the Active Regions (ARs) and lower values in emission in areas where the filaments between ARs and the non-polar coronal holes are located, shown as dark regions close - or inside - the activity belts. The ARs display the typical appearance, similar to those of EUV observations, exhibiting more brightness than the quiet sun. The emission maps in both wavelengths are correlated to the observational data: in Dec. 15 (2001) and Nov. 15 (2003) with EIT/SOHO; in Feb. 05 (2015) with AIA/SDO.

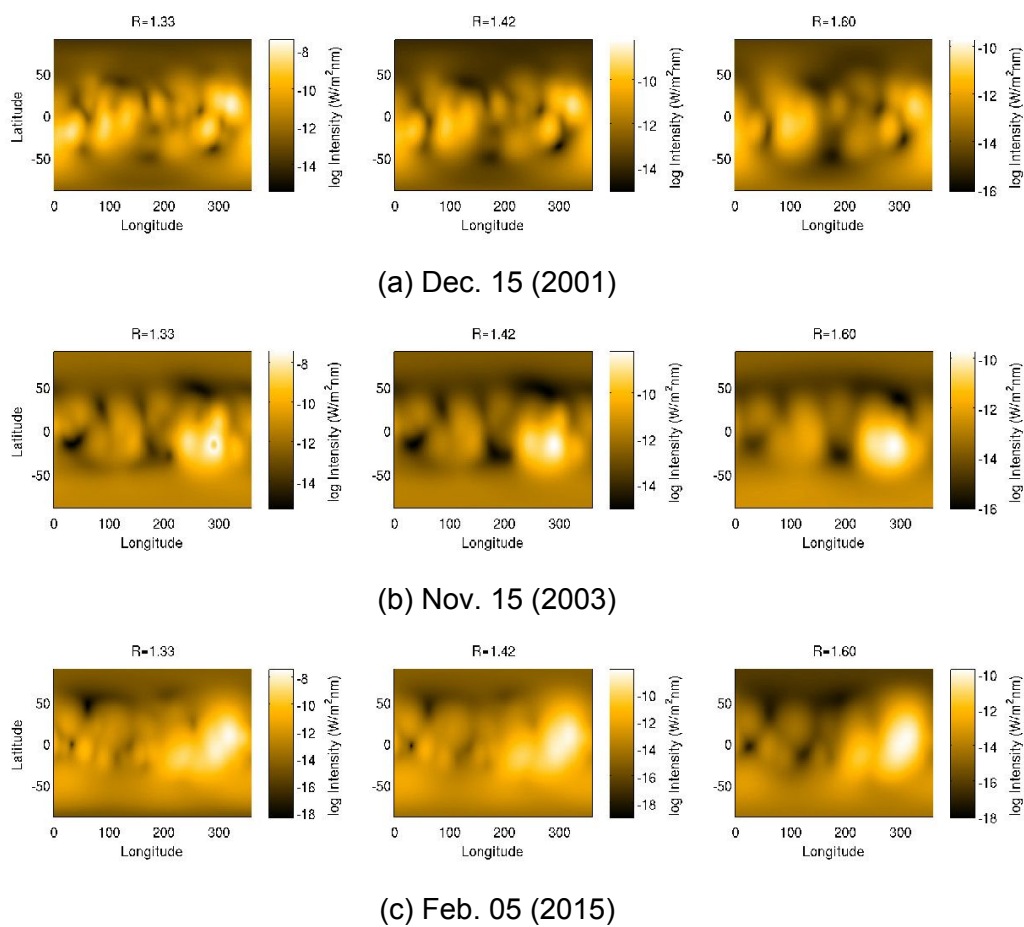

(a) Dec. 15 (2001)

(b) Nov. 15 (2003)

(c) Feb. 05 (2015)

Figure 7: Intensity maps at 19.3nm in three different layers: R = 1.28, 1.43 and 1.60 R. These plots show the different scales in each layer.

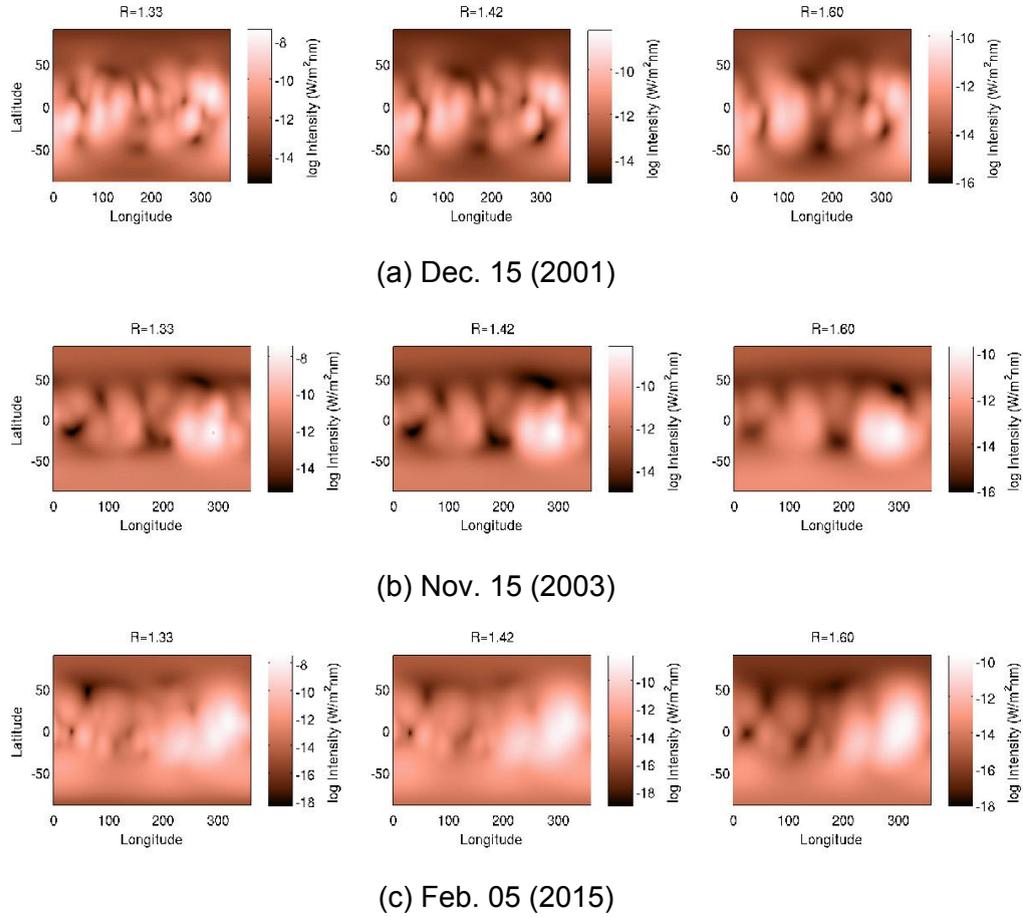

(a) Dec. 15 (2001)

(b) Nov. 15 (2003)

(c) Feb. 05 (2015)

Figure 8: Intensity maps at 21.1nm in three different layers: R = 1.28, 1.43 and 1.60 R. These plots show the different scales in each layer.

## 4 Discussion

The CODET model has not dependence to the filling factors and the areas of the photospheric features, which is a frequent description of the semiempirical models of the Solar Spectral Irradiance (Yeo et al., 2014; Ball et al., 2014; Vieira et al., 2011; Vieira and Solanki, 2010; Krivova et al., 2007; Ermolli et al., 2013). Additionally, the CODET model can describe the evolution of temperature and density in the solar atmosphere using the photospheric magnetic field.

On the other hand, we cannot employ directly the filtergrams from SOHO and HMI instruments in the optimization process, because the computational resources needed are far beyond the computational capability that we have available now. In this way, we decided to present disk-integrated values that could be directly compared to SSI observations to constrain the model. It is important to highlight that our model provides a description of density and temperature through the solar corona based on observations of the solar surface potential magnetic field. As far as we know, it is the first attempt to describe the density and temperature for two solar cycles. We point out that our model, as all models, is an incomplete description of the physical phenomena investigated.

Even the most sophisticated MHD models are employed to describe just some features observed of the evolution of the density and temperature of the solar corona for a very limited time scale. The CODET model employs a general view of the evolution of the magnetic field in the photosphere, expressed in solar synoptic maps. Also, the $B/B_s$ factor tor is dependent on the spatial resolution of the instruments (Chapman et al.,2011), in our case we use data from MDI and HMI; this can contribute to differences between modelled and observational SSI. Also, another reason of these discrepancies can relate to changes during the solar rotation, because of the possibility of unaccounted instrumental drifts (Marchenko et al., 2016). However, using these data sets is possible to reconstruct SSI and TSI (Yeo et al., 2014). Some observational uncertainties can influence the SSI variability, because the uncertainties may differ in units and dimensions from one dataset to the other (Scholl et al., 2016). Besides, the variation in a short time scale was recovered during the solar cycle 23 and 24 (Figure 3 (c) and (d)). In the same way with the result present in Marchenko et al. (2016) for short periods of cycles 23 and 24.

We point out that due to the lack of suitable observations in the EUV spectral region for the period of that we employed to constraint the model, the model parameters were constraint by comparing the model's output to the TIMED/SEE record, which employs model reconstructions to fill data gaps. The temperature and density profiles are strongly dependent on the magnitude of the magnetic field, and two power law exponents describe these variations, α for the temperature and γ for the density profile (Section 2.1). In this description γ have positive values while α have non-positive values in all tests described in Rodríguez Gómez (2017). These power law exponents were found independently through the optimization algorithm Pikaia and it retrieves parameters which describe properly the Solar Spectral Irradiance variations during the last solar cycles. These relationship between the power law exponents generates the temperature profiles that are inversely proportional to the magnetic field, while the density profile is directly proportional to the magnetic field (Figure 5). Other less successful parameter combinations were explored for 17.1nm, 19.3nm, 21.1nm, 33.5nm in Rodríguez Gómez (2017). The temperature profiles show a slight increase in the solar cycle 24 compared with the solar cycle 23. In the external layers the temperature is higher than in the internal layers during the two last solar cycles. The solar cycle 23 shows lower temperature values than the solar cycle 24 due to the modelled temperature profile was described as inversely proportional to the magnetic field (Equation 8, Table 1 and Figure 5-upper panel). Thus, the variations in temperature through the solar atmosphere are highly variable and this behaviour is shown in temperature profiles from CODET model due to the dependence between temperature and emission in the EUV band.

The density profiles show lower values in the external layers and high values in layers near the photosphere. Higher values in the density profiles are more common during the solar cycle 23 compared to the solar cycle 24. The electron density profiles follow the sunspot trend during the solar cycle and they are related to the variations of the magnetic field, because the density profile is proportional to the magnetic field (Equation 5, Table 1 and Figure 5-lower panel). Besides that, the temperature and density profiles in Figure 4 are in accordance to the single-fluid radiative energy balance model through the inner corona presented in Withbroe (1988). That approach employs some empirical values based on measurements of the intensity, polarization of the electron scattered white light corona and measurements of the radial intensity gradients of EUV spectral lines to constraints on temperature in the solar corona. In this model the source of the radiated energy is mechanical energy transported and dissipated in the corona by non-determined mechanisms. The temperature maps obtained from CODET model (Figure 6) show small regions with values in an interval from $\log_{10} T \sim 5$ to $\log_{10} T \sim 6.2$ in the two selected layers. These values were reported in the empirical temperature maps in the EUV wavelengths: 21.1nm, 19.3nm and 17.1nm (Dudok de Wit et al., 2013). Likewise, our results are in agreement to temperatures related to emission measure analysis and with the observations (Aschwanden et al., 2013; Winebarger et al., 2011). The density maps from CODET model are quite well correlated with the magnetic field strength. Tripathi et al. (2008) found that the density in an AR is $10^{10.5}$ and references therein suggests values of $10^9$ cm$^{-3}$, the CODET model shown maximum values of $10^9$ cm$^{-3}$, which is agreement to values reported in the literature. These behaviour are shown in Figure 6 and it is in agreement with the work of Kramar et al. (2014). Additionally, we have focused in the active region belts (Abramenko et al., 2010) and their structures as ARs and non-polar CHs. The emission description (Equation 9) is related to the optically thin emission. Using in the same way to definition of Differential Emission Measure (DEM) (Hannah and Kontar, 2012; Aschwanden, 2005; Warren et al., 1998), in our approach the density and temperature profiles are defined a priori from the equations 5 and 8. Also, we include the contribution function and the atomic data from CHIANTI atomic database 8.0 for EUV lines, in the same way was presented at Fontenla et al. (2014).

The emission maps were explored in three layers: R = 1.33, 1.42, 1.60 R and two wavelengths 19.3nm and 21.1nm. Regions with higher values in emission are related to regions with higher values in density that are the ARs (figures 7, 8 and 6). We mainly focus on ARs and CHs inside a belt of ±40° around the solar equator because the PFSS do not describe regions located in latitudes > ±50° adequately (Abramenko et al., 2010), therefore it is not reliable for modelling polar CHs. This is corroborated by visual inspection of the observed images from EIT/SOHO in 19.5nm and AIA/SDO in 19.3nm and 21.1nm. The ARs and non-polar CHs are reconstructed adequately. Non-polar CHs and regions between ARs that may harbour filaments are also in agreement with observations. Additionally, Warren et al. (2012) describe an inverse correlation between the emission measure (EM) and unsigned magnetic flux for lower (approx, transition region) temperatures, while the emission measure and unsigned flux is directly related for very high temperatures. The inverse EM and unsigned magnetic flux relationship may explain the temperature maps presented in this paper. Therefore the piece of evidence on EUV spectral line considerations and the physics-based model can be in accordance on the results.

## 5 Concluding remarks

The performance of the CODET model is comparable to that of the observational data from TIMED/SEE. The Solar Spectral Irradiance (SSI) variation between solar activity maximum and minimum is properly simulated. The agreement with the data is gratifying considering that the CODET model do not have a MHD approach (Figure 3). Moreover, it is important to highlight the CODET model describe adequately the evolution of magnetic flux from the quiet sun regions or the minimum between the solar cycle 23 and 24 (specifically); this behaviour is probably due to the input of the CODET model. They correspond to the synoptic maps and they are related to the mean variations of the magnetic flux during of a period of ~ 27 days. Then, high-cadence temporal variations in magnetic flux from the active regions are not possible in this current version.

An important feature of the present work is a description of temperature and density profiles in the solar corona in large time scale (two last solar cycles).

Also, temperature, density and emission maps in different layers through the solar atmosphere. The emission is reconstructed for ARs and non-polar CHs in a synoptic view. An interesting relationship between higher values in emission and density are shown and it is expected to be explored in depth in a forthcoming paper.

**Acknowledgements**

This work is partially supported by CNPq/Brazil under the grant agreement no. 140779/2015-9, no. 304209/2014-7 and no. 300596/2017-0. J.P. thanks to the MINECO project AYA2016-80881-P (including AEI/FEDER funds, EU). J.M.R would like to thank Matthieu Kretzschmar and Thierry Dudok de Wit for the helpful discussions during the Sun-Climate Symposium, 2015, in Savannah, Georgia. We thank the anonymous reviewer their comments that helped improving this paper.